\documentclass[aps,twocolumn,superscriptaddress,showpacs,floatfix]{revtex4}
\usepackage{epsfig}
\usepackage{graphicx}
\usepackage{amssymb}

\begin{document}

\title{Effect of bond lifetime on the dynamics of a short-range 
attractive colloidal system}

\author{I. Saika-Voivod}
\affiliation{Dipartimento di Fisica and Istituto Nazionale per la
Fisica della Materia,
Universit\`a di Roma La Sapienza, Piazzale Aldo
Moro~2, I-00185, Roma, Italy}

\author{E. Zaccarelli}
\affiliation{Dipartimento di Fisica and Istituto Nazionale per la
Fisica della Materia,
Universit\`a di Roma La Sapienza, Piazzale Aldo
Moro~2, I-00185, Roma, Italy}
\affiliation{INFM-CRS Soft: Complex Dynamics in Structured Systems,
Universit\`a di Roma La Sapienza, Piazzale Aldo
Moro~2, I-00185, Roma, Italy}


\author{F. Sciortino}
\affiliation{Dipartimento di Fisica and Istituto Nazionale per la
Fisica della Materia,
Universit\`a di Roma La Sapienza, Piazzale Aldo
Moro~2, I-00185, Roma, Italy}
\affiliation{INFM-CRS Soft: Complex Dynamics in Structured Systems,
Universit\`a di Roma La Sapienza, Piazzale Aldo
Moro~2, I-00185, Roma, Italy}

\author{S.~V. Buldyrev}
\affiliation{Center for Polymer Studies and Department of
Physics, Boston University, Boston, MA 02215, USA.}

\author{P. Tartaglia}
\affiliation{Dipartimento di Fisica and Istituto Nazionale per la
Fisica della Materia,
Universit\`a di Roma La Sapienza, Piazzale Aldo
Moro~2, I-00185, Roma, Italy}
\affiliation{INFM-CRS SMC: Center for Statistical Mechanics and
Complexity, Universit\`a di Roma La Sapienza, Piazzale Aldo
Moro~2, I-00185, Roma, Italy}

\date{\today}

\begin{abstract}
We perform molecular dynamics simulations  of short-range attractive
colloid particles modeled by a narrow (3\% of the hard sphere
diameter) square well potential of unit depth.  We compare the
dynamics of systems with the same thermodynamics but different bond
lifetimes, by adding to the square well potential a thin barrier at
the edge of the attractive well.  For permanent bonds, the relaxation
time $\tau$ diverges as the packing fraction $\phi$ approaches a
threshold related to percolation, while for short-lived bonds, the
$\phi$-dependence of $\tau$ is more typical of a glassy system.  At
intermediate bond lifetimes, the $\phi$-dependence of $\tau$ is driven
by percolation at low $\phi$, but then crosses over to glassy behavior
at higher $\phi$.  We also study the wavevector dependence of the
percolation dynamics.
\end{abstract}


\maketitle
\section{Introduction}

Colloidal systems, in which particles are dispersed in a fluid, have
an enormous relevance in industrial applications, owing to the
possibility of chemically or physically tuning the interaction between
the particles and the resulting possibility of designing materials with
novel properties~\cite{bookcolloid,bookhansen,frenkelscience}.   From
the point of view of basic research, colloidal systems are playing a
very important role in  the development of the physics of liquids,
since they open up significantly the range of values of  physically
relevant parameters. For example, novel phenomena arise when the range
of particle-particle interaction becomes significantly smaller than
the size of the particle or when the system is composed of colloidal
particles with significantly different size or mobility.

An interesting phenomenon that is often observed in  colloidal
suspensions, but is absent from atomic or molecular systems, is particle
clustering and gelation.  The gel is  an arrested state of matter at
small values of the packing fraction, a non-ergodic state capable of
supporting  weak stresses. The formal description of gel formation in
colloidal systems has been receiving considerable attention
recently~\cite{fuchs,langmuir,kroy,yukawa,sator}.  Recent numerical
work has also focused on the gelation  problem~\cite{jack,daniel}.
Interesting studies have attempted to provide formal connections
between the formation of a gel and the formation of a glass, both
being disordered arrested states of matter. It is not a coincidence
that such theoretical studies focus on colloidal systems, where
colloid-colloid interaction can be finely tuned, allowing for a
careful test of theoretical predictions.  Indeed, colloids appear to
be ideal systems for unraveling the physics of gel
formation. Understanding the key features of the interaction potential
that  stabilize the gel phase will probably have an impact also on our
understanding of  the protein crystallization
problem~\cite{piazza,caccamo},  where the possibility of generating
crystal structures is hampered by the formation of arrested states.

Sterically stabilized colloidal particles provide an experimental
realization of a system in which the particle-particle interaction can
be well modeled by the hard sphere potential~\cite{vanmegen}. When
this is the case, addition of many small non-adsorbing  polymers
leads, due to depletion mechanisms, to an effective short-range
attraction between the colloidal
spheres~\cite{asakura,likosreview}. Neglecting the effects of the
solvent on the dynamics of the colloidal particles, and integrating
out the behavior of the smaller polymers, one has an experimental
realization of a short-range potential, with a tunable short-range
attraction between particles. The size of the small polymers tunes the
range, while their concentration controls the strength of the
attraction $u_0$.  

At high packing fraction  $\phi \approx 0.6$, these
colloidal systems exhibit the usual hard sphere glassy dynamics.  When
the range of interaction is smaller than about 10 percent of the
hard-sphere diameter, the glass transition line can show re-entrant
behavior~\cite{fuchs,fabbian,dawson,puertasprl,foffipre,zaccarellipre,
pham,bartsh,chen,emanuela}.  That is, in a particular range of $\phi$,
the liquid can be brought to structural arrest by either increasing or
decreasing the ratio $T/u_0$, where $T$ is the
temperature. Experimentally, dynamical arrest phenomena in short-range
attractive colloids are observed not only at high density, as
discussed above, but also in the low packing fraction region.  In this
case, the arrested material is commonly named a gel.  The gel state
displays peculiar features like the appearance of a peak in the static
structure factor, for very large length scales (of the order of
several particle diameters), that is stable in time, as well as a
non-ergodic behavior in the density correlation functions and a finite
shear modulus~\cite{weitz}.  These solid-like, disordered, arrested
features have prompted the appealing conjecture that these colloidal
gels can be viewed as the low-density expression of the high-density
glass line, with both phenomena  being driven by the same underlying
mechanism of arrest~\cite{fuchs,langmuir,kroy,puertaspre}.  However,
such a connection between gelation and the attractive glass is
non-trivial, as pointed out in Ref.~\cite{zaccarelli2001}.  The
presence of an intense pre-peak in the static structure factor  has
suggested also the possibility that, in colloidal systems, the gel
phase is related to a phase separation
process~\cite{bos,jackle,heyes1,heyes2,soga,capri}.  Indeed,  hard
sphere systems with short-range  attraction added tend to phase
separate into a colloid rich phase (liquid) and a colloid poor (gas)
phase. Whether the interaction between  this phase separation and the
re-entrant glass line can bring about a gel phase via arrested phase
separation~\cite{capri} is an idea which is also under current investigation.  
At very low $T$, diffusion limited cluster
aggregation~\cite{bookvicksek,giglio,stprl90X,bibette} may be another way to 
irreversibly obtain a clustered gel (and a frozen pre-peak in
the structure factor).

In the present study, as a step in the process of understanding
gelation in colloidal systems in the absence of phase separation, 
we focus on the dependence of the
dynamics on a purely kinetic factor, the lifetime of the
particle-particle bond.  We introduce a Hamiltonian model of a
short-range attractive colloid, for which we can tune the bond
lifetime, without affecting the thermodynamics.  We have been inspired
by the recent work of Del Gado and coworkers~\cite{delgado}, where a
lattice model was introduced to study the influence of bond lifetime
on the slow dynamics of gelling systems.  Here, we model  a colloidal
system as an ensemble of particles interacting with a short-range
square well, a model sufficiently realistic to properly describe the
physics of short-range systems, but at the same time ideal for
studying particle bonding and percolation since a bond is
unambiguously defined  by the limits of the square well.   In
particular, we study the interplay between percolation and the glass
transition and find that there is a crossover from a percolation
dominated regime, to one controlled by the glass transition.
Differing from Ref.~\cite{delgado}, we find that above the percolation
threshold, changing the lifetime of bonds merely rescales long time
behavior of the system, leaving intact the glassy $\alpha$-relaxation.
Furthermore, we explore the long lifetime case with regards to the
dependence of observing percolation on the wavevector used to probe
the system, making contact with  experimental observables. Finally, as
a contribution towards clarifying the differences between gels and
glasses, we study the same model in the limit of permanent bonds,
where percolation becomes the relevant arrest process in the system.

\section{Molecular Dynamics Simulations - The Model}

We perform collision-driven molecular dynamics simulations of a binary 
mixture of  particles interacting through a narrow  square well pair potential. 
Although colloidal systems are more properly modelled using Brownian dynamics, 
we use molecular dynamics due to its efficiency in the case of step-wise potentials.  
While the short-time dynamics is strongly affected by the choice of the microscopic 
dynamics, the long term structural phenomena, in particular close to dynamical arrest, 
are rather insensitive to the microscopic dynamics~\cite{gleim}.
We use a 50:50 binary mixture of 700 particles of mass $m$
with diameters $\sigma_{AA}=1.2$ and $\sigma_{BB}=1$ (setting the unit
of length).  The hard core repulsion for the $AB$ interaction occurs
at a distance $\sigma_{AB}=(\sigma_{AA} + \sigma_{BB})/2$. The depth of the well
$u_0$ is $1$, and the width $\Delta_{ij}$ of the square well
attraction is such that $\Delta_{ij}/(\sigma_{ij}+\Delta_{ij})=0.03$
for all  interactions between particles of type $i$ and $j$.  $T$ is
measured in units of $u_0$, time $t$ in
$\sigma_{BB}(m/u_0)^{1/2}$. This system has been extensively studied
previously~\cite{zaccarellipre,capri,zacc2,A4}.

\begin{figure}
\hbox
to
\hsize{\epsfxsize=1.0\hsize\hfil\epsfbox{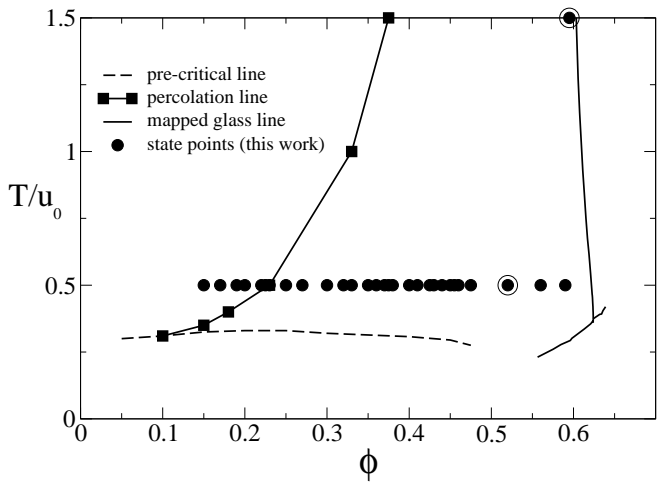}\hfil}
\caption{Phase diagram of the square well binary mixture (reproduced from ~\cite{capri}), 
showing the percolation line (squares), approximate location of the 
spinodal (dashed line), repulsive and attractive glass transition lines 
(solid line). State points studied here are shown as filled circles.  
The highlighted state points 
($T=0.5$, $\phi=0.52$ and $T=1.5$, $\phi=0.595$) refer to those
presented in Fig.~\ref{q20-0.5-0.52} and Fig.~\ref{q20-1.5-0.595},
respectively.}
\label{phasediagram}
\end{figure}

The phase diagram of this system is reproduced from 
Ref.~\cite{capri} in Fig.~\ref{phasediagram}. 
For this model, both the dynamical arrest curves and the spinodal curve have 
been calculated.  The glass line (determined  by extrapolating
the diffusion coefficient calculated in simulation to zero according to a 
power-law~\cite{A4})  shows both a repulsive and an attractive glass 
branch~\cite{zaccarellipre}. The numerical glass lines are well
described by the predictions of mode-coupling theory
(MCT)~\cite{leshouches}, after an appropriate mapping is
performed~\cite{A4,sperl}.  Fig.~\ref{phasediagram} also reports the
static percolation line (defined as the locus of points in ($\phi,T)$
such that 50\% of the configurations possess a spanning, or
percolating, cluster of bonded particles) and the estimated location
of the liquid-gas spinodal (the locus of $T$ below which spinodal
decomposition occurs in simulation). 

It is important to note that in this model the
attractive glass line ends on the spinodal line on the large
$\phi$ branch, proving that arrested states at low $\phi$ in this
model can arise only as a result of interrupted phase
separation~\cite{capri}. It also confirms that, if the MCT predictions for the 
location of the attractive glass are not properly rescaled in the $\phi-T$ plane, 
an incorrect location of the glass line with respect to the spinodal line is predicted.

\begin{figure}
\hbox to\hsize{\epsfxsize=1.0\hsize\hfil\epsfbox{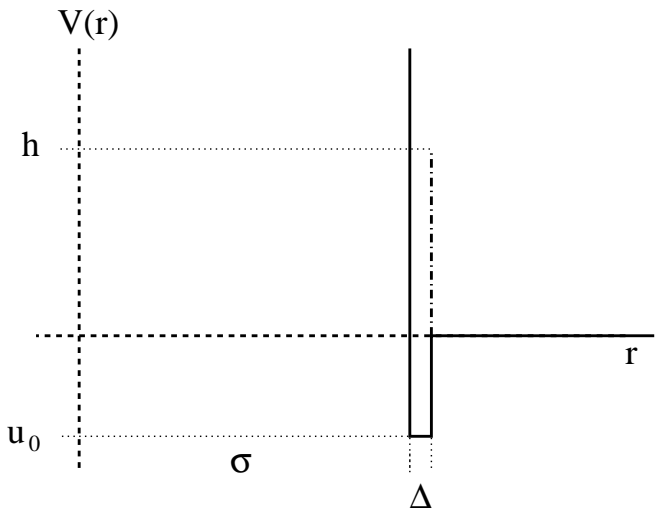}\hfil}
\caption{Schematic of the pair potential.  Here shown are the hard sphere core
diameter $\sigma$, narrow square well of depth $u_0=1$ and width
$\Delta$, with $\Delta/(\Delta+\sigma)=0.03$.  The height $h$ of the barrier
controls the bond lifetime, and hence the microscopic dynamics of the system.
}
\label{potential}
\end{figure}

In order to study the effect of bond lifetime on the dynamics, we add
to the edge of the square well an infinitesimal barrier of tunable
height $h$ (see Fig.~\ref{potential}), thereby stabilizing bonds
formed when particles become trapped in the attractive square well of
the pair potential~\cite{rapaport}. 
As the barrier is infinitesimal, the portion of
phase space occupied is negligible, and hence the thermodynamics  of
the system is unaffected.  For numerical reasons, in the
code we have implemented a  barrier width of  $3 \times 10^{-4}
\sigma_{BB}$ checking that  the static structure of the system is not
affected  by this tiny but non-zero width.

To the extent that the thermodynamics are
unaffected by the barrier,  configurations drawn from equilibrium
simulations of the $h=0$ case are also representative configurations
of the system when $h\neq0$.  Thus, 
results for different $h>0$ are obtained using $30$ or more
independent initial configurations obtained by equilibrating the 
system for $h=0$.  This technique alleviates the
computational burden when working with large values of $h$.

In this article we focus on the time dependence of the
(collective) density-density correlation function (dynamic structure
factor).  The dynamic structure factor, the
correlation function typically accessed in scattering experiments, is
given by  $F_q(t) \equiv  \left< \rho_q(t)\rho_{-q}(0)\right>/S(q)$,
where $\rho_q(t)=\frac{1}{\sqrt{N}} \sum_{i=1}^{N} \exp{(-i\vec
q\cdot \vec r_i)}$, 
$S(q)=\left< \left|\rho_q(0)\right| \right>^2$ is
the static structure factor,  $\left< . \right>$ denotes an ensemble
average, $\vec r_i$ is the position vector of a particle,
$\vec q$ is a wavevector and $i$ labels the $N$ particles of the system.
We also make use of the correlation function for type $A$ particles
only, defined similarly as
$F_q^A(t) \equiv  \left< \rho^A_q(t)\rho^A_{-q}(0)\right>/S^A(q)$,
where
$\rho^A_q(t)=\frac{1}{\sqrt{N_A}} \sum_{i=1}^{N_A} \exp{(-i\vec
q\cdot \vec r_i)}$, $S^A(q)=\left< \left|\rho^A_q(0)\right| \right>^2$
is the partial static structure factor for type $A$ particles,
and the summations are over the $N_A$ particles of type $A$.
The qualitative behavior of $F_q(t)$ and $F_q^A(t)$ is the same for
this system.

\section{The infinite bond lifetime case}

We start discussing the case of bonds of  infinite lifetime, i.e., the
case where $h \rightarrow \infty$. In this limit, a well defined model
for continuum percolation is generated. The spatial distribution of
the particles is fully determined by the equilibrium properties of the
square well potential (and hence static correlations are known and
precisely defined) while the dynamics is the dynamics of a system
constrained by irreversible bonds.  The possibility of generating
equilibrium structures with $h=0$ to be used as starting
configurations for the case $h \ne 0$ allows us to completely decouple
issues arising from the bond lifetime from issues  associated with
non-equilibrium properties and aging also when the packing fraction is
larger than the percolation value.  The averaging over different
starting configurations allows us to properly sample  configuration
space. To study the $h \rightarrow \infty$ case we perform simulations
at $T/u_0=0.5$, for about 30 different values of packing fraction, as
indicated in Fig.~\ref{phasediagram}.  Numerically, we achieve the
infinite limit by setting $h=1000$, a value high enough so that we
never observe bond breaking.

\begin{figure}
\hbox to\hsize{\epsfxsize=1.0\hsize\hfil\epsfbox{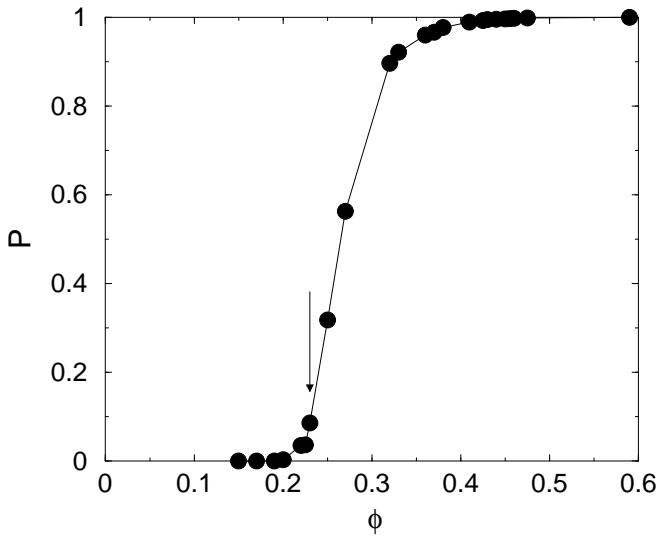}\hfil}
\caption{$P$ as a function of $\phi$.  The arrow indicates the
$\phi_p=0.23$ at which 50\% of the sampled configurations contain a percolating
cluster.}
\label{pinf}
\end{figure}

\begin{figure}
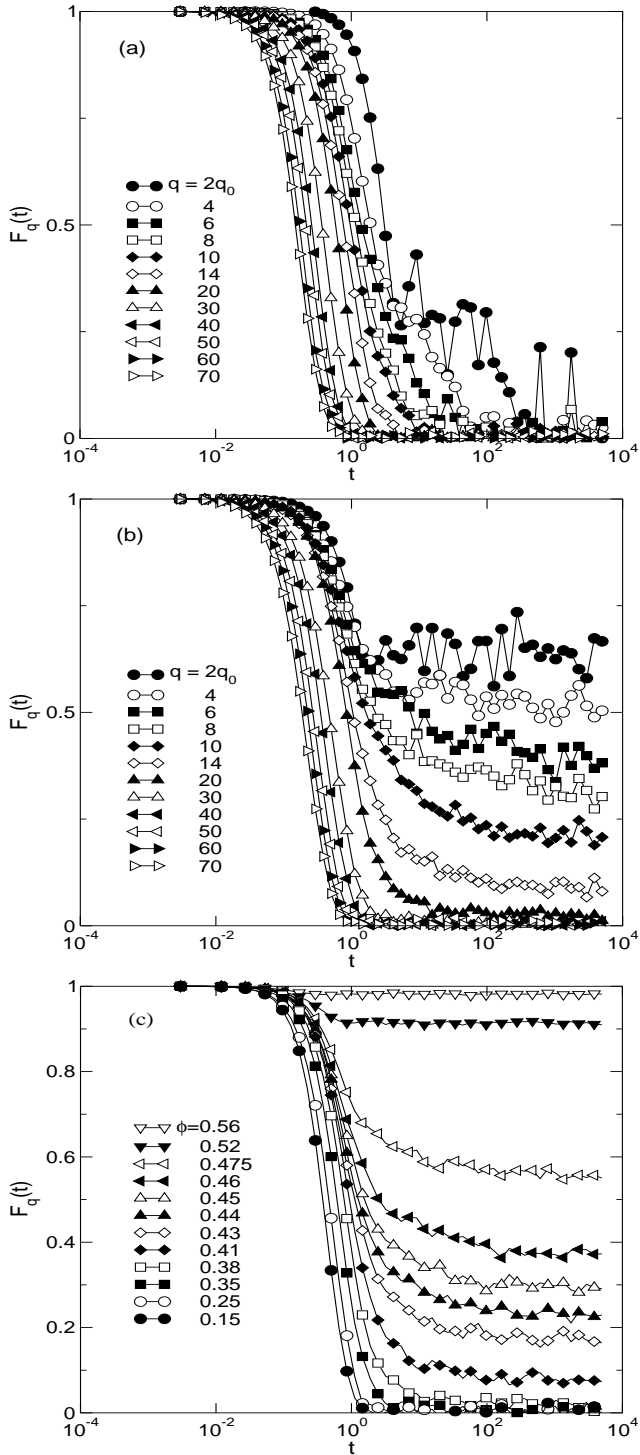

\includegraphics[width=3.3in,height=2.5in]{fig4a.eps}
\vskip 0.1cm
\includegraphics[width=3.3in,height=2.5in]{fig4b.eps}
\vskip 0.1cm
\includegraphics[width=3.3in,height=2.5in]{fig4c.eps}
\caption{Dependence of $F_q(t)$ on $\phi$ and $q$ for $h=\infty$.
(a) $F_q(t)$ at $\phi=0.225$ (just below the percolation threshold of $0.23$).
Here, $q\sigma_{BB}=nq_0$ for various $n$ shown in the legend, 
with $q_0=\pi/L=0.2408$ ($L$ is the length of the simulation box).  
At $n=2$, it is more difficult to
reduce noise in the data because of the small number of $q$-vectors
available for averaging.
(b) $F_q(t)$ at $\phi=0.38$ (well above percolation), with $q_0=0.2867$.
(c) $F_q(t)$ at $q\sigma_{BB}\approx 2\pi$ for various $\phi$.   
From such curves we determine the plateau height $f_q$.}
\label{fig:ftqh1000}
\end{figure}

The $\phi$ dependence of $P(\phi$), the fraction of particles
belonging to the spanning cluster, provides a way of detecting the
location of the percolation point. 
In all percolated configurations, we observe the presence of only one
spanning cluster.
When finite size effects are
negligible, $P\sim |\phi-\phi_p|^{\beta}$ where $\beta$ is a critical
exponent~\cite{staufferbook,torquatobook}.   Fig.~\ref{pinf} shows
$P(\phi)$.  The arrow in the figure indicates $\phi_p=0.23$, 
which we identify as the value
of the packing fraction at which a spanning cluster is  found in 50
percent of the configurations~\cite{miller}.   To estimate the effect
of bonding on the dynamics, we show in Fig.~\ref{fig:ftqh1000} the
packing fraction and wavevector dependence of $F_q(t)$.

For $\phi < \phi_p$ (Fig.~\ref{fig:ftqh1000}(a)), correlation
functions decay to zero, independently of the value of $q$, as
expected for a system where only diffusive clusters of finite size are 
present. For $\phi > \phi_p$  (Fig.~\ref{fig:ftqh1000}(b)), an
``infinite'' spanning cluster is present. On increasing $\phi$ the
size (mass) of the spanning cluster increases progressively,
incorporating  90\% of the particles in the system already when $\phi = 0.32$ 
(see Fig.~\ref{pinf}).  For $\phi > \phi_p$,  wavevectors are characterized by 
correlation functions which do not decay to zero any longer,  
reflecting the presence of a non-relaxing component of the density fluctuations.

\begin{figure}
\hbox to\hsize{\epsfxsize=1.0\hsize\hfil\epsfbox{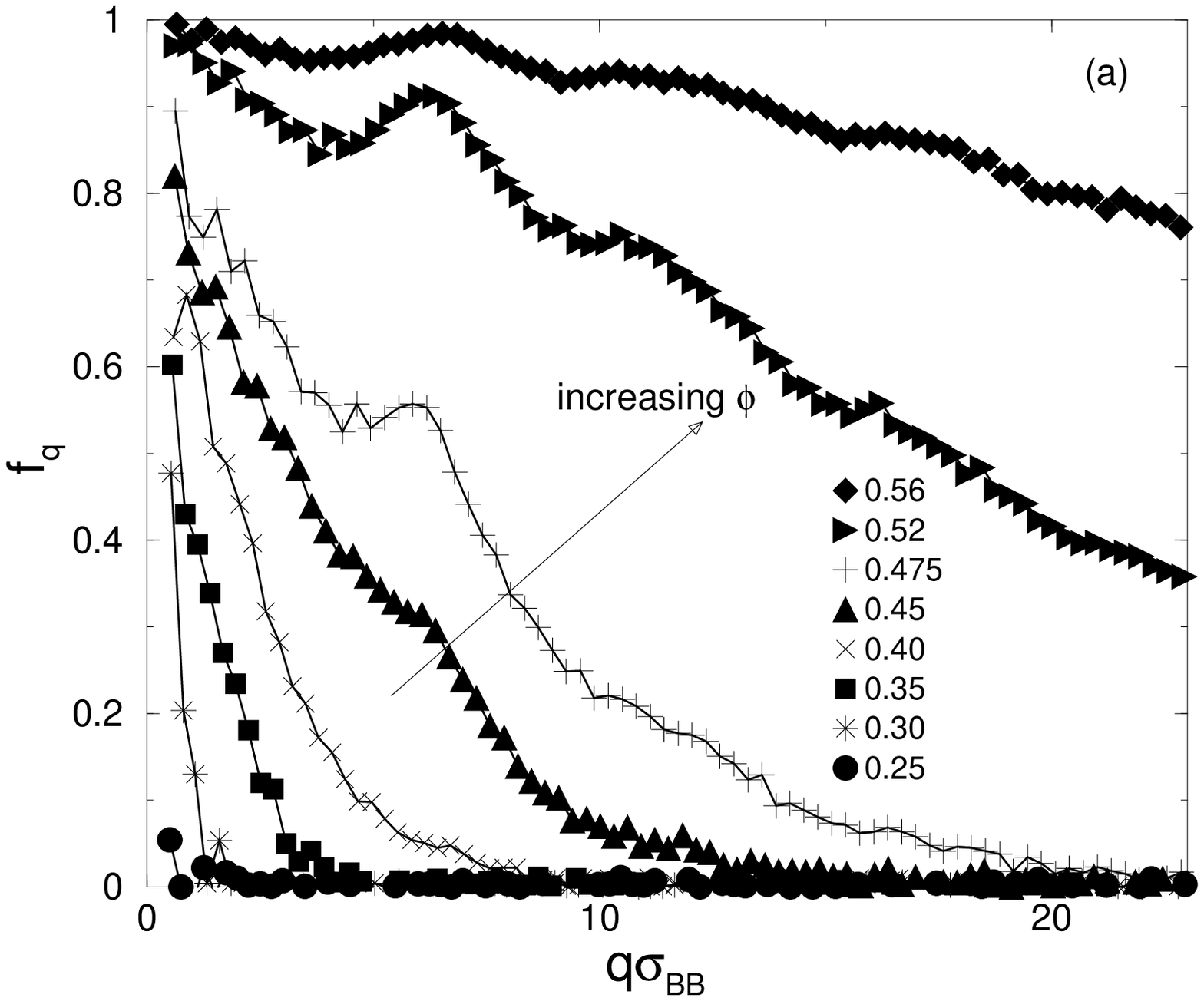}\hfil}
\hbox to\hsize{\epsfxsize=1.0\hsize\hfil\epsfbox{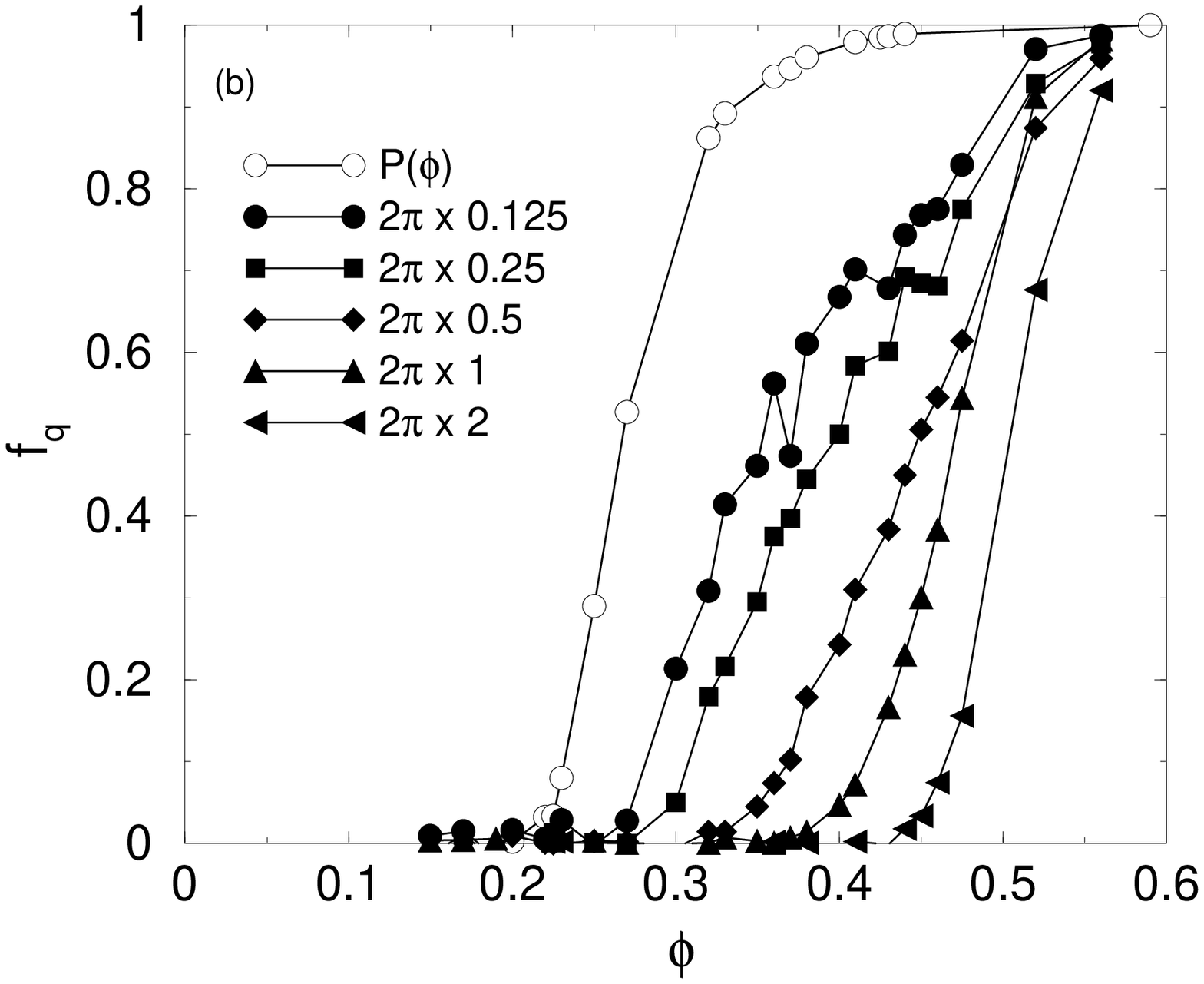}\hfil}
\caption{(a) Plateau height $f_q$ as a function of $q$ for various $\phi$,
for infinite $h$ (irreversible permanent bonds).  Legend gives $\phi$
values.
(b) Connection to percolation: $f_q$ as a function of $\phi$ for
various $q$.  The legend indicates values of $q\sigma_{BB}$ for
$f_q$ curves corresponding to filled symbols. The curve labeled 
with open circles shows $P$, the average fraction of  particles 
participating in a percolating cluster (taken from Fig.~\ref{pinf}).}
\label{p_vs_q}
\end{figure}

Close to percolation, only for  very small
wavevectors does $F_q(t)$ go to a non-zero plateau of height $f_q$, also
called the non-ergodicity factor. On increasing packing fraction,  the
amplitude of the plateau increases significantly, as shown in
Fig.~\ref{fig:ftqh1000}(c).  Simultaneously, correlation functions at
larger and larger wavevectors show a finite non-zero long time limit.
The wavevector and $\phi$ dependence of $f_q$
is shown in Fig.~\ref{p_vs_q}.

The appearance of a non-zero $f_q$, whose amplitude and width grow
on increasing $\phi$ is consistent with the onset of a percolation transition 
and the loss of ergodicity of the particles in the infinite spanning cluster.  
The inverse of the half-width of $f_q$ provides an estimate of the associated 
localization length. On increasing $\phi$ beyond percolation,  such a length 
decreases from infinity (or from the simulation box length in a finite size 
system)  down to the dimension of the particles, in analogy with the progressive 
decrease of the connectness length of the spanning 
cluster~\cite{torquatobook,chiarificazione}.    Owing to the large localization 
length close to percolation and to the  tenuous structure of the percolating 
cluster, close to percolation   a non-zero $f_q$ value can be clearly detected only 
at very small wavevectors. On increasing $\phi$, the increase in the number of 
particles in the infinite cluster and the associated decrease of the length 
generate an increase in the amplitude and width of $f_q$, making possible the 
numerical observation of a non-zero $f_q$ even at large $q$.

It is interesting to compare the behavior of the non-ergodicity factor
observed in the case of percolation with the case of the glass.  
The most striking difference is in the change of $f_q$ across the glass and 
percolation transitions. In the case of glasses, $f_q$ shows a discontinuous jump, 
while in the case of percolation it increases  from zero continuously.  
In the language of MCT~\cite{lorentzgas}, the percolation transition is 
analogous to what is 
called a  ``type A'' transition while the ordinary glass transition is of ``type B''.  

Another important aspect is the fact that the width in $q$ space
of $f_q$ (e.g., the $q$ value at which a curve in 
Fig.~\ref{p_vs_q}(a)
reaches half its height) is of the order of the inverse of the
nearest neighbor distance in the case of glasses (or even larger in
the case of attractive glasses) while it is extremely small close to
percolation. Only when most of the particles are part of the spanning
cluster, does
the width of $f_q$ become compatible with the one of
glasses. This change in the width of $f_q$
reflects the significant difference in
localization length of the particles
(the length scale on which particles are trapped in chiefly
vibrational motion) at the glass transition  (of the
order of the  nearest neighbor or of the bond distance) as opposed to
the very large localization length at the percolation transition where
a tenuous (almost massless in the thermodynamic limit) spanning
cluster appears.

It is important to note that in the case of glasses, the glass
transition is marked by the arrest of density fluctuations on every
length larger than the nearest neighbor distance, while in the case of
percolation, the observation of a non-ergodic transition is strongly 
dependent on the observation length. To clarify this point,
Fig.~\ref{p_vs_q}(b)
shows the $\phi$ dependence of $f_q$ at several $q$ values (i.e. a cut
of the data shown in Fig.~\ref{p_vs_q}(a)
at fixed $q$).  We note that
the steep increase of $f_q$ from zero occurs at larger and larger
$\phi$ values on increasing $q$.  This suggests that experiments --- capable 
of measuring a non-zero $f_q$ with a finite precision --- restricted to a fixed $q$ value
will notice a  loss of ergodicity in the sample, as reflected by a non-zero long 
time limit of the correlation function, only at a $\phi$ value which may 
be significantly larger than the percolation packing fraction.  

In analogy to the connection between $\phi_p$ and the vanishing  of  $P(\phi)$ 
(also shown in Fig~\ref{p_vs_q}(b)), one may define an apparent 
$\phi_c(q)$ based on the packing fraction at which the $f_q$ curves shown in 
Fig.~\ref{p_vs_q}(b) cross a fixed value, controlled by the precision of the 
experimental technique in detecting a non-zero $f_q$ value.
This $\phi_c(q)$ could be considered an indicator of the percolation transition as  
observed at a particular $q$ vector.


\section{Finite bond lifetime: Effect of barrier height on $F_q(t)$}

\begin{figure}
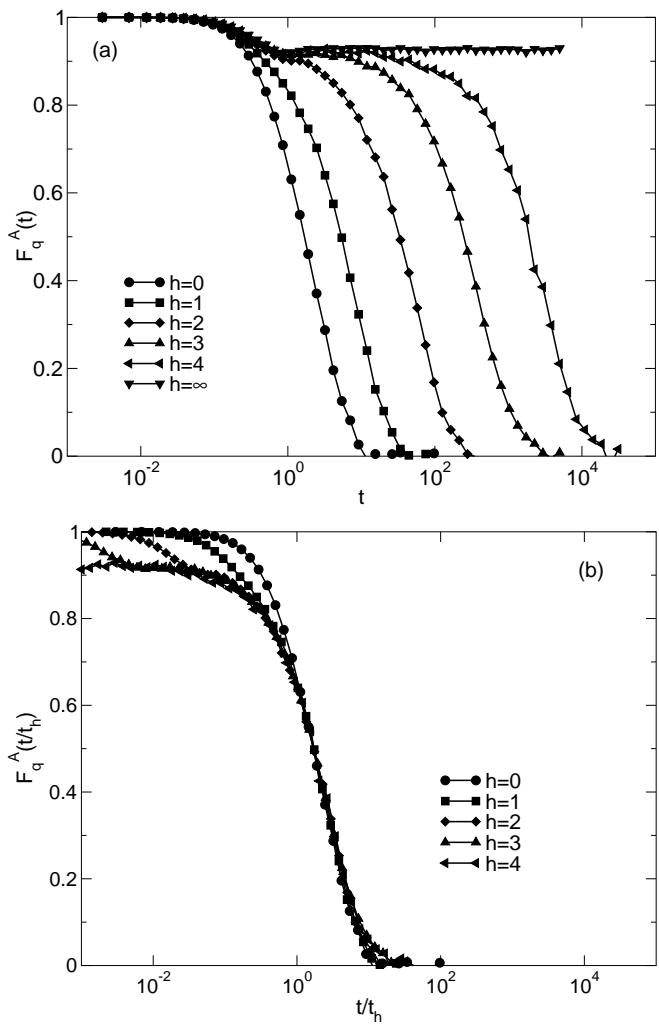

\hbox to\hsize {\epsfxsize=1.0\hsize\hfil\epsfbox{fig6a.eps}\hfil}
\vskip 0.2cm
\hbox to\hsize {\epsfxsize=1.0\hsize\hfil\epsfbox{fig6b.eps}\hfil}
\caption{(a) The dynamic structure factor at $q\sigma_{BB}=6.37$ for
the state point $\phi=0.52$, $T=0.5$ (see Fig.~\ref{phasediagram}) for
various values of the barrier height $h$.  The attractive plateau
persists longer as $h$ increases, i.e., as the bond lifetime is
increased.  Panel (b) shows that the functions collapse onto a common
curve at long times, when $t$ is rescaled (see text).}
\label{q20-0.5-0.52}
\end{figure}

When $h$ has a finite value, the lifetime of the bond is
finite. Hence, a new time scale enters into the description of
dynamics in the model. In particular, we are interested in the
modification of the density correlation functions introduced by the
finite bond lifetime, and in the competition between the bond  time
scale and the caging time scale close to the glass transition.  In
Fig.~\ref{q20-0.5-0.52}(a) we plot $F_q^A(t)$ at $\phi=0.52$  and
$T=0.5$ for $q\sigma_{BB} \approx 2 \pi$, for several increasing
values of $h$. When $h=0$, the decay of the correlation function does
not show signatures of two-step relaxation, owing to the location of
the state point in the re-entrant liquid portion of the phase diagram
(state point is highlighted in
Fig.~\ref{phasediagram})~\cite{zaccarellipre}.   As $h$ increases, two
new features appear: (i) a slowing down of the correlation function
and (ii) the emergence of a two step-relaxation process. Correlation
functions decay to a rather high plateau before decaying to zero at
long times.   When $h=\infty$, the correlation function does not decay
to zero any longer.  In Fig.~\ref{q20-0.5-0.52}(b),
despite the emergence of a
plateau, the long time behavior  is merely rescaled with respect to
the $h=0$ case. The shape of the correlation functions does not change
significantly in the long time regime and indeed  curves for different
(finite) $h$ values can be superimposed on the $h=0$ curve, as shown
in Fig.~\ref{q20-0.5-0.52}(b). 
As a suitable scaling time  we choose the time at which the
correlation functions reach $20\%$ of their initial value.
More precisely,  the density correlation functions are
plotted as a function of a rescaled time,
$t/t_h$, where $t_h = \tau_{20}(h)/\tau_{20}(0)$,
and $F_q^A(\tau_{20}(h)) = 0.20$.

\begin{figure}
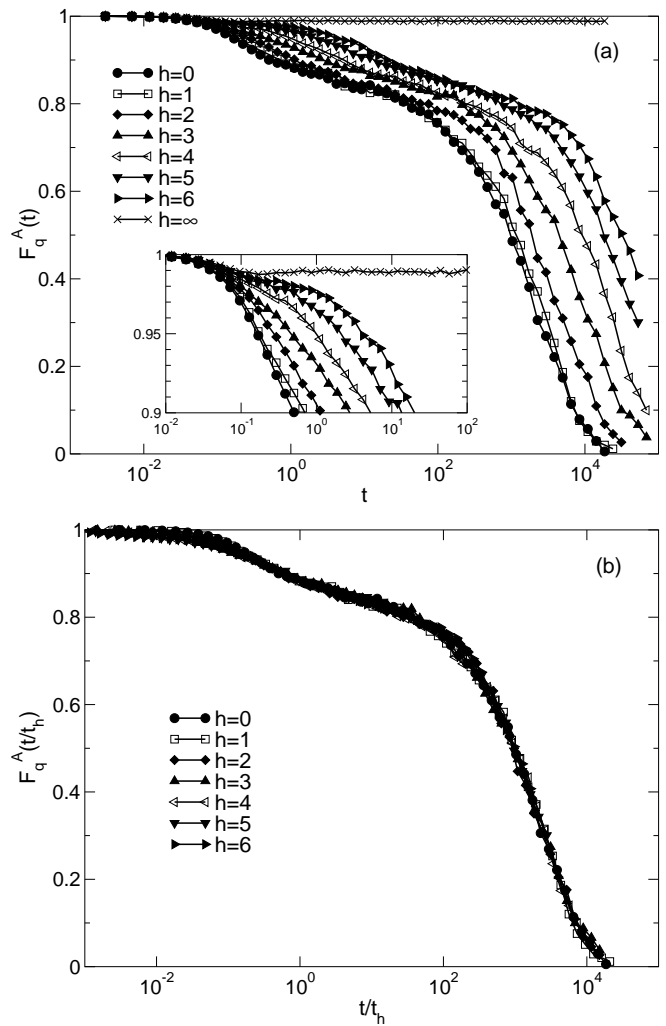

\hbox to\hsize{\epsfxsize=1.0\hsize\hfil\epsfbox{fig7a.eps}\hfil}
\vskip 0.2cm 
\hbox to\hsize{\epsfxsize=1.0\hsize\hfil\epsfbox{fig7b.eps}\hfil}
\caption{(a) The dynamic structure factor at $q\sigma_{BB}=6.67$ for
the state point $\phi=0.595$, $T=1.5$ (see Fig.~\ref{phasediagram})
for various values of the
barrier height $h$.  (b) We again see the stabilization of the attractive
plateau (highlighted in the inset), and simple time rescaling at long
times.}
\label{q20-1.5-0.595}
\end{figure}

We next discuss the case in which, in the absence of a barrier, the
liquid is close to a repulsive glass transition, marked by the presence of a
two-step relaxation in $F_q^A(t)$.  We show in
Fig.~\ref{q20-1.5-0.595}(a) the 
correlation functions for different barrier
heights for the state point $T=1.5$, $\phi=0.595$  (highlighted in
Fig.~\ref{phasediagram}).  At this higher $T$, the $h=0$ system
behaves as a hard sphere binary mixture. The slow dynamics is thus
characteristic of repulsive glass dynamics and shows a well defined
plateau.   On increasing $h$,  one observes a progressive modification
of the  correlation function at short times, and the emergence of the
``gel'' plateau (highlighted in the inset).  When time becomes
comparable to the bond lifetime, the correlation function leaves the
``gel'' plateau and approaches the caging plateau, following the same
dynamics as in the $h=0$ case, as clearly indicated by the
superposition of curves for different $h$ values on a rescaling of the
time (Fig.~\ref{q20-1.5-0.595}(b)). This superposition indicates that
the lifetime of the bond indeed acts to renormalize the ``microscopic
time''.  An increase in $h$ increases the time required for breaking
the particle-particle bonds which in turn increases the timescale of
the breaking and reforming of cages.

The fact that the slow relaxational processes are unaffected by the
bond lifetime (apart from a trivial scaling factor depending on $h$)
is particularly reassuring for theories of the glass transition which
connect static properties to dynamic properties, like MCT. The
peculiarity of this model is indeed the fact that static properties
are independent of $h$. Hence, according to MCT all dynamical
properties associated with caging should be independent of $h$.  The
scaling observed in Fig.~\ref{q20-0.5-0.52}(b) and
Fig.~\ref{q20-1.5-0.595}(b) supports  such a hypothesis.  We also
note that the results reported here differ from those  reported in
Ref.~\cite{delgado}, where clustering induced by the bonds was
considered to be significantly connected to the glass transition
phenomenon. One possible explanation of such a  difference may lie in
the fact that in the study of Ref.~\cite{delgado}, at odds with the
present model, the bond lifetime is strongly coupled to the structure
of the system.

\section{Crossover from percolation to glassy dynamics}

\begin{figure}
\hbox to\hsize{\epsfxsize=1.0\hsize\hfil\epsfbox{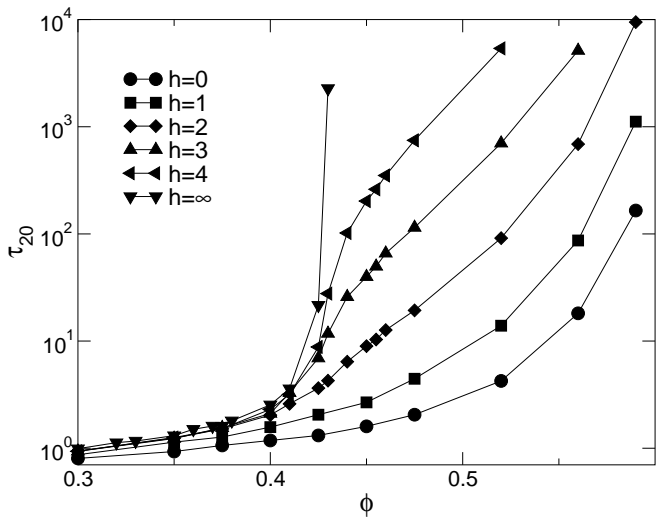}\hfil}
\caption{The dependence of $\tau_{20}$ on $h$ and $\phi$, obtained
from $F_q^A(t)$ at $q\sigma_{BB}=2\pi$.   The curves are for $h=0$
(circles), $h=1$ (squares), $h=2$ (diamonds), $h=3$ (up-triangles),
$h=4$ (left-triangles) and permanent bonds (down-triangles).  For
permanent bonds, $\tau_{20}$ diverges near $\phi=0.43$.  For smaller
values of $h$, $\tau_{20}$ at smaller $\phi$ tracks the divergence at
$\phi=0.43$, but then crosses over to glassy dynamics with a
divergence at higher $\phi$.}
\label{q20-tau-phi}
\end{figure}

We now focus on the $\phi$ dependence of the characteristic time for
different values of $h$.  To quantify the characteristic time, we
use  $\tau_{20}$, the time at which $F_q^A(t)$ decays to a value
of $0.2$.  In Fig.~\ref{q20-tau-phi} we show the dependence of
$\tau_{20}$ on $h$  and $\phi$ for wavevector modulus
$q=2\pi/\sigma_{BB}$.

We see from Fig.~\ref{q20-tau-phi} (and from Fig.~\ref{phasediagram})
that $\tau_{20}$ for the $h=0$ system appears to diverge near
$\phi \gtrsim 0.6$.  For the case of  permanent bonds, $h=\infty$,
$\tau_{20}$ instead shows an apparent divergence at $\phi_c\approx0.43$.
As discussed in the previous sections,  this divergence is a manifestation 
of the percolation transition which has taken place at $\phi_p= 0.23$
(see Fig.~\ref{p_vs_q}(b), near $\phi\approx0.43$ for
$q\sigma_{BB}=2\pi$).  At
intermediate values of $h$, the $\phi$ dependence of $\tau_{20}$
is highly non-trivial.  A  crossover from the
percolation behavior ($h=\infty$) to the glass behavior ($h=0$) takes
place at a typical time controlled by the value of $h$. This is most
clearly seen for $h=3$ and $h=4$ in Fig.~\ref{q20-tau-phi}.  Indeed,
deviations from the $h=\infty$ case are expected  when the lifetime of
the bond becomes shorter than $\tau_{20}(h=\infty)$. In another way,
on time scales shorter than the bond lifetime, the bonds appear to be
permanent and the system behaves like the $h=\infty$ case.

\begin{figure}
\vskip 0.5cm
\hbox to\hsize{\epsfxsize=1.0\hsize\hfil\epsfbox{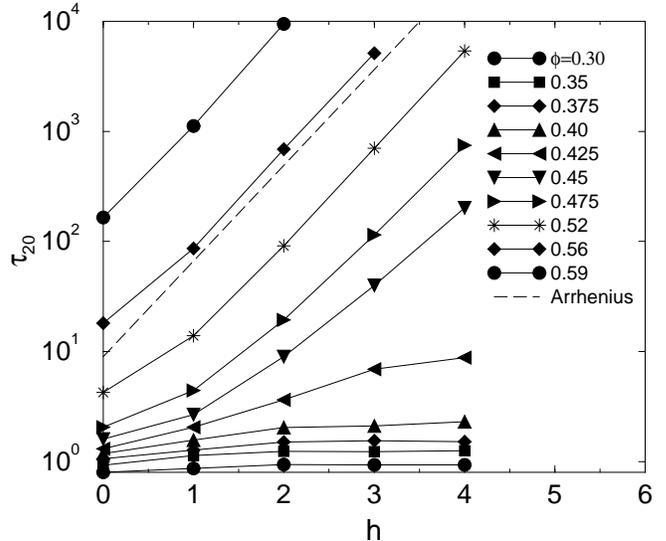}\hfil}
\caption{The dependence of $\tau_{20}$ on $h$ for various packing
fractions, at $q\sigma_{BB}=2\pi$. Above $\phi_c$,   the
characteristic time tends to an Arrhenius behavior with $h$, i.e.,
$\tau_{20}\approx \exp{(h/T)}$.}
\label{q20-tau-b}
\end{figure}

To estimate the role of $h$ in slowing down the dynamics we report
$\tau_{20}$ as a function of $h$ for various isochores in Fig.~\ref{q20-tau-b}.
We find that above $\phi_c$, $\tau_{20}$ approaches an
Arrhenius behavior with respect to $h$ as $h$ increases, i.e.,
$\tau_{20}(\phi,h) \approx g(\phi) \exp{(h/T)}$ where $g(\phi)$ is
function only of the state point chosen.  This factorization of time
allows us to clarify that the main effect of bonding is to  redefine
the microscopic time scale of the dynamics. The bond lifetime does not affect  
the properties of the ($\alpha$-relaxation) slow dynamics on approaching the  
glass transition.

It is interesting to state the connection of the present findings with an earlier 
work~\cite{zacc2}, where the barrier was used to extend the high-plateau to timescales 
associated with $\alpha$-relaxation of the native system, in the packing fraction region 
where a glass-glass transition was expected. The present data clearly show that the 
barrier does not affect (except for a rescaling of the microscopic time) the true 
$\alpha$-relaxation dynamics. On the other hand, large barrier values bring into the 
window of experimental observation the intra-well dynamics and move to inaccessible 
regions the $\alpha$-relaxation.  Under these conditions, the decay of the correlation 
function in the experimentally accessible window is limited to the high-plateau $f_q$ value, 
which coincides with the attractive glass.

This suggests the possibility of simultaneously studying gelation and
glassy dynamics within the same system, by focusing on different time
scales.  For example, in a system with transient bonds, observation on
time scales much shorter than the bond lifetime
would reflect the percolation dynamics, and hence
gelation, while observation on time scales much longer than the bond
lifetime would yield results driven by glassy dynamics.

\section{Conclusions}

In this paper we have introduced a simple model for studying continuum
percolation in a system with well defined spatial correlations between
the particles. In this model, bonding ambiguities are suppressed by
the square well shape of the potential.  While studying this model
with permanent bonds, we have focused on the behavior of the
density-density correlation function across the percolation
transition, defined as the packing fraction at which a spanning
cluster appears. We have found that the behavior of the density
fluctuations is significantly different from the one characteristic of
supercooled liquids and glasses. In the percolation problem, the
non-ergodicity parameter increases  continuously from zero and the
localization length is extremely large and becomes comparable to the
one observed in glasses (of the order of about 0.1 of the nearest
neighbor or less) only well inside percolation.  In this respect,
percolation (with infinite lifetime bonds) and the glass transition
are two distinct phenomena with distinct experimentally detectable
signatures.  We have also shown that in the case of percolation,
since the range of wavevectors where non-ergodic behavior is observed
grows with $q$ on increasing $\phi$ (for $\phi > \phi_p$), experiments
at fixed wavevector (due to their intrinsic finite resoluation)
detect a non-ergodic transition at a packing
fraction larger than $\phi_p$. In the case of glasses, the
observation of an ergodic to non-ergodic behavior is essentially
identical at all $q$ values (except at very large $q$, describing self
intra-cage motion~\cite{highq}).

The model allows us also to study the effect of the finite bond
lifetime while altering neither the structure nor the thermodynamics
of the system.  A comparison at different bond lifetimes is thus
performed on configurations which are characterized by the same
particle-particle correlation.  This study confirms the results
recently reported by Del Gado and coworkers~\cite{delgado} for a
lattice model concerning the existence of a crossover in the dynamical
properties from a percolation controlled dynamics to a glassy dynamics
on increasing $\phi$, when the lifetime of the bond is longer than the
microscopic particle dynamics in the absence of a barrier.  However,
our results differ from  Ref.~\cite{delgado} in that  we find that the
bond lifetime acts essentially as a redefinition of the microscopic
time and does not  alter any feature of the slow dynamics and of the
scaling laws approaching the glass transition. Still, the dynamics at
times shorter than the  $\alpha$-relaxation time  is strongly affected
by the finite lifetime of the bond.  
The addition of the barrier, which increases the bond lifetime, extends 
the duration of the  plateau characteristic of short-range attractive glasses
(Fig.~\ref{q20-1.5-0.595}(a) inset).  Since here we are in the liquid regime,
the duration of the high-plateau is controlled by the
(tunable) bond lifetime. For times longer than
$\exp{(h/T)}$ the correlation function leaves this plateau and
approaches the (lower) plateau associated with caging dynamics
(Fig.~\ref{q20-1.5-0.595}(a)), and
then finally relaxes to zero --- leaving the intrinsic slow long time
dynamics of the system intact.

\section{Acknowledgments}
We thank Emanuela Del Gado for helpful
discussions.
We acknowledge support from MIUR Cofin 2002, Firb and 
MRTN-CT-2003-504712.
I.~S.-V. acknowledges NSERC (Canada) for funding
and SHARCNET for computing resources.


\begin{thebibliography}{999}
\bibitem{chiarificazione} 
In this respect, as a first approximation, 
$f_q$ should be considered to be related to the Fourier transform of the spatial 
correlation in the infinite cluster.
\bibitem{highq}
A similar apparent shift in the location of the glass transition
could be observed in the case of glasses if $q \gg 2\pi/\sigma$.
\end{thebibliography}
\end{document}